\documentclass[12pt,preprint]{emulateapj}
\usepackage{apjfonts}
\usepackage[dvipdfm]{hyperref}
\usepackage{natbib}
\usepackage{apjfonts}
\begin{document}

\title{Quenching depends on morphologies: implications from the ultraviolet-optical radial color distributions in Green Valley Galaxies}
\shortauthors{Pan et al.}
\shorttitle{UV-optical colors in green galaxies}
\author{Zhizheng Pan\altaffilmark{1}, Jinrong Li\altaffilmark{2,3}, Weipeng Lin \altaffilmark{1}, Jing Wang\altaffilmark{4}, Xu Kong\altaffilmark{2,3} }

\email{panzz@shao.ac.cn, linwp@shao.ac.cn}

\altaffiltext{1}{Key laboratory for research in galaxies and cosmology, Shanghai Astronomical Observatory,
Chinese Academy of Science, 80 Nandan Road, Shanghai, 200030, China}

\altaffiltext{2}{Center of Astrophysics, University of
Science and Technology of China, Jinzhai Road 96, Hefei 230026, China}
\altaffiltext{3}{Key Laboratory for Research in Galaxies and Cosmology, USTC,
CAS, China}
\altaffiltext{4}{CSIRO Astronomy \& Space Science, Australia Telescope National Facility, PO Box 76, Epping, NSW 1710, Australia}

\begin{abstract}
In this Letter, we analyse the radial UV-optical color distributions in a sample of low redshift green valley (GV) galaxies, with the \emph{Galaxy Evolution Explorer} (\emph{GALEX})+Sloan Digital Sky Survey (SDSS) images, to investigate how the residual recent star formation distribute in these galaxies. We find that the dust-corrected $u-r$ colors of early-type galaxies (ETGs) are flat out to $R_{90}$, while the colors turn blue monotonously when $r>0.5R_{50}$ for late-type galaxies (LTGs). More than a half of the ETGs are blue-cored and have remarkable positive NUV$-r$ color gradients, suggesting that their star formation are centrally concentrated; the rest have flat color distributions out to $R_{90}$. The centrally concentrated star formation activity in a large portion of ETGs is confirmed by the SDSS spectroscopy, showing that $\sim$50\% ETGs have EW(H$\rm \alpha$)$>6.0$ \AA. For the LTGs, 95\% of them show uniform radial color profiles, which can be interpreted as a red bulge plus an extended blue disk. The links between the two kinds of ETGs, e.g., those objects having remarkable "blue-cored" and those having flat color gradients, are less known and require future investigations. It is suggested that the LTGs follow a general picture that quenching first occur in the core regions, and then finally extend to the rest of the galaxy. Our results can be re-examined and have important implications for the IFU surveys, such as MaNGA and SAMI.

\end{abstract}

\section{Introduction}
Why and how galaxies stop their star formation activities and then move from the "blue cloud" to the "red sequence" (the so-called "quenching procedure") is a key question in the studies of galaxy formation and evolution. In literature, at least two factors are proposed to affect quenching: galaxy stellar mass and environmental conditions \cite[e.g.,][]{Peng 2010}. However, it is still far from a comprehensive understanding of the detailed quenching picture.

In the color-magnitude (or the color-mass diagram), the green valley is a narrow region connecting the "blue cloud" and "red sequence" \citep{Strateva 2001, Baldry 2004,Bell 2004}. The galaxies in GV were thought to be the transition populations between the star-forming and the quenched galaxies \citep{Bell 2004, Wyder 2007, Mendez 2011, Pan 2013}, hence holding important clues to the consequence led by the quenching procedure. In a recent paper, \citet{Schawinski 2014} showed that the color distributions of GV ETGs on the $u-r$ versus. $NUV-u$ diagram can be modeled by a quenching time scale of $\tau_{\rm quench} \sim 0.1$ Gyr, while for the bulk of LTGs a longer quenching time scale is required ($\tau_{\rm quench} \sim 2.5$ Gyr). This study suggests that morphology is another factor which can affect the detailed quenching processes in galaxies.

Recent works have found that quenching is connected with remarkable bulge growth \citep{Bell 2012, Chueng 2012, Pan 2013,Bruck 2014}, regardless of its detail working mechanism. It is thus necessary to investigate in which part the quenching procedure first take place in a galaxy. Recently, \citet{Abramson 2014} found the specific star formation (sSFR) re-normalized by disk stellar mass (formulated by $\rm sSFR_{\rm disk}\equiv SFR/M_{\ast,disk}$) are weakly dependent on disc masses. In their work, "mass quenching" was interpreted as more massive galaxies have larger bulge mass fractions$-$the portion of a galaxy not forming stars. Thus when investigating how quenching is processing in galaxies, it will be more reasonable to treat galaxies as composite systems, rather than integrated mass, if the data allows to do so.

Spatial resolved study on the residual star formation in a GV galaxy may shed light on how quenching is processing in a galaxy. Large surveys such as Sloan Digital Sky Survey (SDSS) \citep{York 2000} and \emph{Galaxy Evolution Explorer (GALEX)} \citep{Martin 2005} survey have provided a large sample of local galaxies to facilitate such kind of studies. In a previous work, \citet{Suh 2010} have investigated the $g-r$ color profile in a sample of low redshift ETGs and found that roughly 30\% of them show positive color gradients, which is consistent with the existence of central star formation. Since the emissions in the UV bands are more sensitive to recent star formation than that of the optical bands, in this Letter, we use \emph{GALEX}+SDSS data to study the radial UV-optical color distributions in a sample of low redshift face on GV galaxies, and to investigate how the residual star formation activities distribute in transition galaxies. Throughout this Letter, we assume a concordance $\Lambda$CDM cosmology with $\Omega_{\rm m}=0.3$, $\Omega_{\rm \Lambda}=0.7$,
$H_{\rm 0}=70$ $\rm km~s^{-1}$ Mpc$^{-1}$.

\begin{figure*}
\centering
\includegraphics[width=160mm,angle=0]{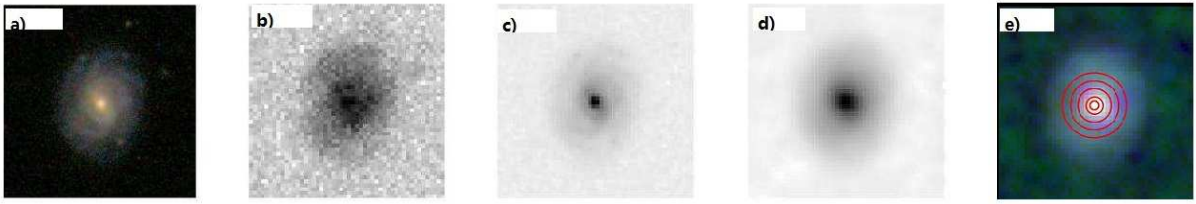}
\caption{We select a LTG from our sample and show its images. From left to right, the images are: a) SDSS gri color image; b) NUV image; c) registered SDSS r-band image, interpolated to the \emph{GALEX} resolution; d) image in c) convolved NUV PSF; e) RGB color image, generated by NUV (blue), SDSS u-band (green) and SDSS r-band (red) PSF-matched images. The stamp sizes are 72$\times$72 \arcsec. The red circles represent 5 photometric apertures with r$=[1.5, 3.0, 6.0, 9.0, 12.0]\arcsec$, respectively.}
\end{figure*}

\section{Sample selection}
Our parent sample is drawn from \citet{Schawinski 2014}, which contains $\sim$ 46,000 galaxies at the redshift range of $z=[0.02,0.05]$. This sample is magnitude completed to $M_{\rm z,Petro}=-19.5$ AB mag and with Galaxy Zoo \citep{Lintott 2008,Lintott 2011} visual morphological classifications \footnote{The sample is downloaded form http://data.galaxyzoo.org/.}. The stellar masses are derived by fitting the five SDSS photometric bands to a library of 6.8$\times10^6$ models star formation histories generated from \citet{Maraston 1998,Maraston 2005} stellar models. We follow the process of \citet{Schawinski 2014} to select GV galaxies. First the galaxies are k-corrected to $z=0$ using the KCORRECT code of \citet {Blanton 2007} with the SDSS five broad-band photometry. Then the magnitudes are corrected for dust reddening using estimates of internal extinction from the stellar continuum fits by \citet {Oh 2011}, applying the \citet {Cardelli 1989} law. Then the GV is defined on the dust-corrected color-mass diagram for all galaxies, given
\begin{equation}
-0.70+0.25\times M_{\ast}< ^{0.0}(u-r)<-0.30+0.25\times M_{\ast}
\end{equation}
, where $^{0.0}(u-r)$ is the k-corrected $u-r$ color index and $M_{\ast}$ is log stellar mass, respectively. This criterion is stricter than that of \citet{Schawinski 2014} to ensure that our samples are not contaminated by the blue cloud and red sequence galaxies. With this criterion, about 700 ETGs and 3000 LTGs are classified as GV galaxies. To ensure that their radial color distributions can be robustly measured, only galaxies with $b/a>0.7$ are selected. The remaining sample contains about 300 ETGs and 700 LTGs.

\section{Data reduction}

The \emph{GALEX} NUV image has a resolution of 1 pixel=$1\arcsec. 5$ and a point spread function (PSF) with full width at half-maximum (FWHM)=$5\arcsec. 3$. The pixel size and PSF of SDSS image are $0\arcsec .396$ and $1\arcsec .4$, respectively. To ensure that the photometric measurements are consistent across the different bands, we need to transform all the images to the same geometry and effective resolution before doing photometry.

Following \citet{Wang 2010}'s pipeline, a UV-optical matched photometric catalog have been generated by Li et al. (2014, in prep). This catalog contains about 220,000 galaxies with uniform photometric measurements on the resolution and PSF-matched \emph{GALEX}+SDSS images. Here we briefly describe our data reduction. We first cross-matched the SDSS DR8 \citet{Aih11} spectroscopic galaxies with the \emph{GALEX} GR6 frames. Then we downloaded the matched galaxy images form the SDSS and \emph{GALEX} database. For the \emph{GALEX} images, Only those with exposure times larger than 1000 seconds are used. We degraded the pixel scales of the SDSS images and registered the frame geometries to the corresponding NUV images. To match the \emph{GALEX} and SDSS photometry, the SDSS image needs to be convolved the NUV PSF kernel function. The PSF kernel function was obtained by fitting a two-dimensional Gaussian function to the stacked star stamps in the corresponding \emph{GALEX} NUV frame. The PSF-convolving procedure on the SDSS images was done by running \textsc{IMMATCH/PSFMATCH} task in \textsc{IRAF}\footnote{http://iraf.noao.edu/}. Finally, aperture photometry were done by running \textsc{SExtractor} \citep{Bertin 1996} on the resolution and PSF-matched \emph{GALEX} and SDSS images, over 5 different apertures, with r=[1.5, 3.0, 6.0, 9.0, 12.0] \arcsec. The \textsc{SExtractor} also measures the AUTO magnitude, e.g., the total magnitude of a galaxy. Fig. 1 shows an example of registered and PSF-convolved images for a LTG in our sample.

The PSF-matched photometry have been compared with the SDSS public data. Good consistency is found between the SDSS c-model magnitudes and the AUTO magnitudes we measured. Finally, the magnitudes are corrected for galactic extinction using the galactic dust map \citep{Schlegel 1998}.

We cross-match the GV sample selected in Section 2 with the UV-optical matched catalog and obtain 117 ETGs and 219 LTGs. Among these, 9 ETGs and 13 LTGs classified as Seyfert galaxies on the BPT diagram are rejected \citep{Baldwin 1981}. The final sample containing 108 ETGs and 206 LTGs will be used to investigate their NUV-optical radial color distributions in the next section.

\section{Results}
\subsection{u-r and NUV-r radial color distributions}

In Fig. 2 we show the radial dust-corrected SDSS $u-r$ color distributions. The dust correction on the colors is done by using a single E(B-V) value (drawn from \citet{Oh 2011}) at all radii, as it is difficult to know how extinction vary at different radii. The radii are normalized by $R_{50}$, the radii enclosed 50\% SDSS r-band petrosian fluxes. To present the results more clearly, we divide the sample into two subsamples, according to their $C_{\rm r=1\arcsec .5}-C_{\rm auto}$. $C_{\rm r=1\arcsec .5}$ and $C_{\rm auto}$ are the color indices measured in the central $r=1\arcsec .5$ aperture and that of the whole galaxy, respectively. A galaxy having $C_{\rm r=1\arcsec .5}-C_{\rm auto}<0.0$ means that its color in the core is bluer than that of the whole. We call galaxies of this category "blue-cored". In this sense, "red-cored" galaxies have $C_{\rm r=1\arcsec .5}-C_{\rm auto}>0.0$.
\begin{figure}
\centering
\includegraphics[width=80mm,angle=0]{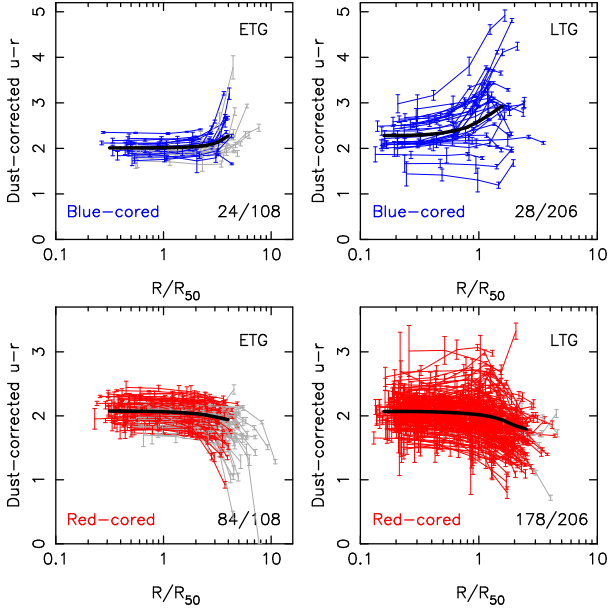}
\caption{The radial dust-corrected $u-r$ color profiles. The Left panels show the results of ETGs, and LTGs are shown in right panels. To be clarified, we binned the samples into two subsamples according to their $C_{\rm r=1\arcsec .5}$-$C_{\rm auto}$. The radial color profiles s of blue-cored galaxies are shown in upper panels (blue lines). The grey lines show the color distributions with SDSS $R_{\rm 90}$$<8.0$ arcsec. The thick black line shows the median color profile. }
\end{figure}

As also shown in \citet{Schawinski 2014}, the GV ETGs are averagely less massive than LTGs. ETGs also have more compact morphologies and approximately 35\% of our ETGs have $R_{90}<8.\arcsec 0 $. So the measurements of the outer colors in these small-size ETGs will be seriously contaminated by the background and not reliable. For the $R_{90}<8.\arcsec 0$ galaxies, we show their color profiles in grey lines. Note that rejecting these small-size galaxies will not affect our main conclusions on the radial color distributions. However they will be kept to maintain a large ETG sample size.

24 out of 108 ETGs are blue-cored, as shown in the upper left panel in Fig. 2. This fraction is significantly larger than that of LTGs, which is only 28/206. For the red-cored subsamples, we find their color profiles are different between ETGs and LTGs: most ETGs show very mild color gradient out to 3$R_{50}$ ($\approx R_{90}$ for ETGs), whereas LTGs colors decrease when r>0.5$R_{50}$.

Fig. 3 shows the radial NUV$-r$ color distributions. NUV$-r$ is a widely explored color index in literature \citep{Salim 2005, Wyder 2007, Wang 2010}. Interestingly, this figure shows dramatically different features compared to the $u-r$ color distributions. For ETGs, we find 68/108 of them having inner-blue NUV$-r$ colors, e.g. their recent star formation are concentrated in the central regions. The rest show very flat NUV$-r$ color profiles.

For LTGs, more than 90\% of them show uniform NUV$-r$ color distributions. Their NUV$-r$ colors monotonously decrease beyond r$\sim$0.5$R_{50}$, which can be interpreted by red bulges plus blue disk components that are still actively forming stars. We also find that the integrated NUV$-r$ colors of ETGs are systematically redder than those of LTGs, confirming the findings of \citet{Schawinski 2014}. Thanks to the \emph{GALEX} photometry, we are able to investigate the recent star formation activities in GV galaxies in more detail. It is impossible when only the SDSS photometry is available.

\begin{figure}
\centering
\includegraphics[width=80mm,angle=0]{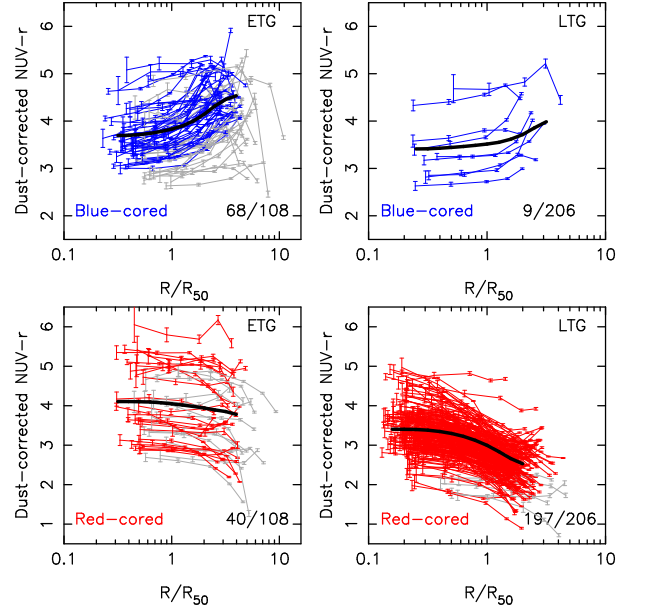}
\caption{Same as Figure.2 but shown in dust-corrected $NUV-r$ colors. }
\end{figure}

\begin{figure*}
\centering
\includegraphics[width=160mm,angle=0]{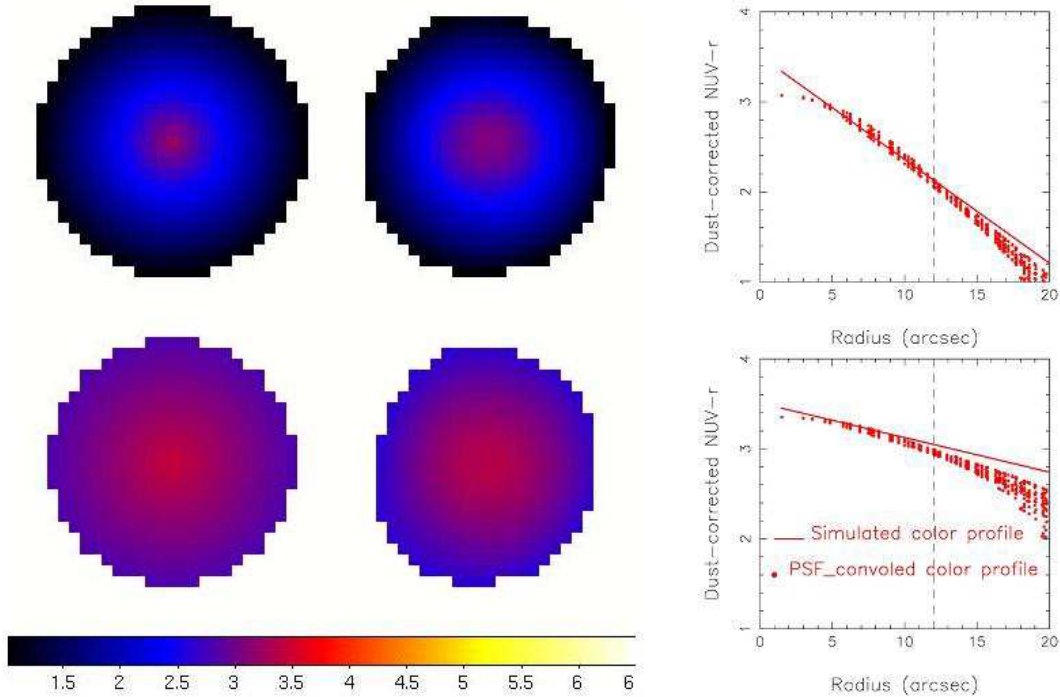}
\caption{In the top panels we simulate the galaxy to have a steep NUV-$r$ color gradient. The left panel show the NUV-$r$ color map measured on the images without convolving the NUV PSF. The one shown in the middle panel is measured on the PSF-convolved images. The right panel shows the comparison between the two kinds of color profiles. The thick red lines indicates the given color profile. The red dots indicate the color index measure on each pixel. the vertical line indicates our color profile detecting threshold, $r=12\arcsec.0$. In the lower panels, we simulate the galaxy to have a mild gradient.  }
\end{figure*}

\subsection{A test on the color profile on simulated images}
The issue we most concerned is how the PSF impacts on the galactic color profiles. Is the color profile we measured on an image with a $5\arcsec.3$ PSF robust?

To answer this question, we have done a test on the simulated images. For each galaxy, we use its SDSS r-band image to simulate its corresponding image in the NUV band. To be simplified, here we only discuss the simplest case. First the SDSS r-band image is degraded to $1\arcsec.5$/pixel to match the NUV resolution. The NUV flux at the radius $r$ of the simulated image ($f_{\rm NUV,r}$) is then given:
\begin{equation}
f_{\rm NUV,r}=f_{\rm SDSS~r-band,r}\times 10^{kr}\times 10^{-3.5/2.5}
\end{equation}
, where $f_{\rm SDSS~r-band,r}$ is the flux at $r$ on the SDSS r-band image and $k$ is a given gradient value, respectively. By definition, The NUV$-r$ color at $r$ is:
\begin{equation}
(\rm NUV-\emph{r})_{\rm r}=2.5 \rm log_{10}(f_{\rm SDSS~r-band,r}/f_{\rm NUV,r})=3.5-2.5kr
\end{equation}
. The value $(\rm NUV-\emph{r})_{\rm r=0}$=3.5 is the median color at galactic centers, which is directly taken from Fig. 3. Another set of image, namely the simulated PSF-convolved image, are also created by convolving the image we simulated to the NUV PSF. Finally we measure the NUV$-r$ color index on the two sets of simulated images to investigate how the PSF will impact on the output color profile.

We select a LTG with $R_{90}= 15\arcsec.0$ and show the result in Fig. 4.  In the upper and lower panels, we simulate a color profile with a steep and a mild gradient, respectively. Interestingly, we find the shape of output color profile looks quite similar to those measured on the real images. For both cases, we find that the color profiles measured on the simulated PSF-convolved image \emph{are basically consistent} with the prior given profiles. The color profile is smoothed at galactic centers by the PSF, as seen in Fig. 4. We find that the color deviation becomes evident at $r>15\arcsec.0$, which is likely owing to the low surface brightness at the radii larger than $R_{90}$. To conclude, we find the color profile for a $R_{90}\sim 10\arcsec.0$ galaxy is \emph{not seriously} modified by the NUV PSF, and our result are of high robustness.

\subsection{Star formation activities at galaxy centers}
\begin{figure}
\centering
\includegraphics[width=80mm,angle=0]{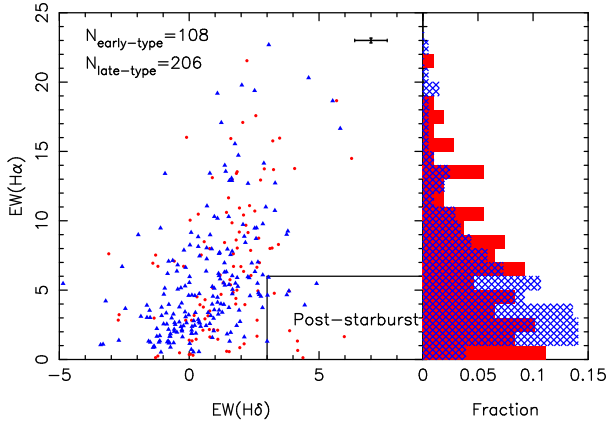}
\caption{Left: The EW(H$\delta$) vs. EW(H$\rm \alpha$). diagram. ETGs are shown in red solid circles and LTGs are shown in blue triagles. The mesurements are drawn from the MPA catalog. Right: The EW(H$\rm \alpha$) distributions. ETGs are shown in solid red histogram and LTGs are in shadow blue histogram, respectively. The box encloses galaxies defined as E+As. }
\end{figure}

The H$\rm \alpha$ emission is a good tracer of instantaneous SFR \citep{Kennicutt 1998}. In this subsection, we investigate the star formation in the central 3\arcsec.0 region in our sample using the SDSS spectroscopy.  The EW(H$\rm \alpha$) values are taken from the MPA/JHU catalog\footnote{http://www.mpa-garching.mpg.de/SDSS/DR7}. In Fig. 5 we plot the EW(H$\rm \delta$) versus. EW(H$\rm \alpha$) diagram. It can be seen that the bulk of GV galaxies have weak H$\rm \alpha$ emissions, i.e., EW(H$\alpha$)<10 \AA, showing that the star formation rate is low in the core of GV galaxies. In the subsection above we have found that a large fraction of ETGs have blue-cored NUV$-r$ colors. Interestingly, In Fig. 5 we find that nearly half of ETGs have EW(H$\rm \alpha$)>6.0 \AA, confirming that they have star formation in the core regions. We call galaxies with EW(H$\rm \alpha$)>6.0 \AA\ as "H$\rm \alpha$ emitter" in the following. This fraction is significantly lower in LTGs, indicating that the residual star formation in LTGs are mostly attributable to the galactic disc.

The star formation strength was found to link with galaxy mass, as demonstrated by the famous star-forming "main sequence" \citep{Peng 2010, Fang 2012}. We have check whether the different core EW(H$\alpha$) distributions in ETGs and LTGs are due to their different mass distributions. The sample is divided into two subsamples according to their stellar mass. The dividing mass threshold is set to log$(M_{\ast}/M_{\sun})$=10.2, the mass that can roughly separate galatic properties into two main groups \citep{Kauffmann 2003,Nair 2010}. The H$\rm \alpha$ emitter fraction is $21/39\approx0.54$ and $34/69\approx0.49$ in the low and high mass ETG subsamples, respectively. The fractions is $28/64\approx0.44$ and $43/142\approx0.30$ in the LTGs, respectively. Thus the different EW(\rm H$\rm \alpha$) distributions between ETGs and LTGs can not be explained by their different mass distributions.

Under the $\Lambda$CDM paradigm, an ETG is formed through hierarchical mergers. The E+A galaxies, also called post-starburst galaxies, are considered as the products of galaxy interactions and the progenitors of red ETGs. In Fig. 4, the box enclosed EW(H$\delta$)>3.0 \AA\ and EW(H$\alpha$)<6.0 \AA\ defines our E+A selection criterion. It can be seen that most E+A galaxies have early-type morphologies, although the number of LTGs are significantly larger than ETGs. However, even in GV ETGs, E+A galaxies only compose a rather small fraction, suggesting that most red ETGs may not undergo the E+A phases in their past.

\section{Summary and discussion}

Our main finding is that GV ETGs have dramatically different radial NUV$-r$ color distributions compared to LTGs. A larger fraction of ETGs have blue cores, both represented by $u-r$ color and NUV$-r$ color. More than a half of ETGs show blue-cored NUV$-r$ colors and the remainders have flat color distribution out to $R_{90}$. Compared to ETGs, nearly all LTGs have uniform color profiles that can be well interpreted as a red bulge plus a blue disk component. The more fraction of central star formation in ETGs than in LTGs are confirmed by the SDSS spectra. We will discuss our findings in this section.

\subsection{Implications on ETGs evolution}

The inner-blue phenomenon in ETGs is not first reported by us but have been mentioned in some previous works \citep{Jeong 2009, Suh 2010}. It is likely attributable to a starburst or its residual in the galactic center. Under the $\Lambda$CDM paradigm, a major merger of two gas rich LTGs will induce gas inflow to trigger central starburst \citep{Springel 2005}. Strong starburst will soon be suppressed by AGN feedback and the star formation shut down quickly, then produce an E+A remnant, as predicted by models \citep{Hopkins 2006}. Interestingly, \citet{Schawinski 2010} found roughly 50\% blue ETGs show merger features in deep images, supporting this scenario. Some previous studies also found a large fraction of E+A galaxies show very obvious interaction features \citep{Goto 2005,Yang 2008}. However, the E+A population only compose a very small portion of the total GV ETGs. We suggest most red ETGs may not undergo the E+A phases in their past.

The star formation in the cores of ETGs can be alternatively explained by the contribution of a "cooling flow" \citep{Brown 1998,Wang 2010}. In the X-ray observations, some ETGs have relatively lower temperature in the core regions and can induce gas inflow. The inflow of cool gas can fuel an AGN or trigger star formation activity in the center of ETGs. This picture explains the star formation in the center without the need of a strong starburst triggered by a merger.

\subsection{Implications on LTGs evolution}
We find most GV LTGs host inactive cores and blue disk components. The central masses of LTGs can be built either by secular evolution \citep {Kormendy 2004} or the central starbursts. \citet{Wang 2012} found that LTGs with strong bars either have enhanced central star formation rates, or star formation that is suppressed compared to the mean.  \citet{Masters 2011} found that redder LTGs are more likely to have bars. These findings suggest that the central mass build up and the subsequent quench in a portion of LTGs is likely driven by bars. Other mechanisms such as galaxy interactions are needed to explain the quenched LTGs without bars.

The color gradient of LTGs is consistent with inside-out disk growth scenario. \citet{Wang 2011} found evidence that disk galaxies follow an inside-out disk growth model. In this model, the HI gas are accreted onto the outer disc and form new stars, through which the disk galaxy grow up. The long quenching time scale required for a LTG as modeled in \citet{Schawinski 2014} is likely due to the long time scale of removing/exhausting the cool gas on its gaseous disk.

However, we note that several LTGs have inner-blue color profiles. We have check their SDSS images and found, 4/9 of them have visible tidal interaction features. Among these, one is classified as an E+A galaxy. Thus the central star formation in these LTGs is likely driven by galaxy interactions. The rest have much smaller bulge components than most LTGs and their SDSS spectra show relative strong H$\rm \alpha$ emission lines. Thus they are possibly in the process of bulge building. However, the fraction of blue-cored LTGs is so small and will not hamper our final conclusion.

\acknowledgments
This work was supported by the NSFC projects (Grant Nos. 11121062, 11233005, U1331201, 11225315, 11320101002), the Specialized Research Fund for the Doctoral Program of Higher Education (SRFDP, No. 20123402110037), and the ``Strategic Priority Research Program the Emergence of Cosmological Structures'' of the Chinese Academy of Sciences (Grant No. XDB09000000, XDB09010000).


\begin{thebibliography}{99}
\bibitem[\protect\citeauthoryear{Abramson et al.}{2014}]{Abramson 2014}Abramson
K., et al., 2014, ApJ, 785L, 36A
\bibitem[\protect\citeauthoryear{Aihara et al.}{2011}]{Aih11} Aihara, H., et al., 2011, apjS, 193, 29
\bibitem[\protect\citeauthoryear{Baldry et al.}{2004}]{Baldry 2004}Baldry, I.
K., et al., 2004, ApJ, 600, 681
\bibitem[\protect\citeauthoryear{Baldwin et al.}{1981}]{Baldwin 1981}Baldwin, J.
A, Philips, M. M. \& Terlevich R. 1981, PASP, 93, 5
\bibitem[\protect\citeauthoryear{Bell et al.}{2004}]{Bell 2004}Bell, E. L, et
al., 2004, ApJ, 608, 752
\bibitem[\protect\citeauthoryear{Bell et al.}{2012}]{Bell 2012}Bell, E. L, et
al., 2012, ApJ, 753, 167
\bibitem[\protect\citeauthoryear{Bertin \& Arnouts}{1996}]{Bertin 1996}Bertin E. \& Arnouts S., 1996, A\&AS, 117, 393
\bibitem[\protect\citeauthoryear{Blanton \& Roweis}{2007}]{Blanton 2007}Blanton M.
R., Roweis, S. 2007, AJ, 133, 734
\bibitem[\protect\citeauthoryear{Bruck et al.}{2014}]{Bruck 2014}Bruck, A. F. et al., 2014, MNRAS, 441, 599B
\bibitem[\protect\citeauthoryear{Brown \& Bregman}{1998}]{Brown 1998}Brown, B. A. \& Bregman, J. N, ApJL, 495L, 75B
\bibitem[\protect\citeauthoryear{Cardelli, Clayton \& Mathis}{1989}]{Cardelli 1989}Cardelli, J. A., Clayton, G. C., Mathis, J, S., 1989, ApJ, 345, 245

\bibitem[\protect\citeauthoryear{Chueng et al.}{2012}]{Chueng
2012}Chueng, E, et al., 2012, ApJ, 760, 131C
\bibitem[\protect\citeauthoryear{Fang et al }{2012}]{Fang
2012}Fang, G, Kong, X, Chen, Y, Lin, X, 2012, ApJ, 751, 109F
\bibitem[\protect\citeauthoryear{Goto }{2005}]{Goto
2005}Goto, T., 2005, MNRAS, 357, 937
\bibitem[\protect\citeauthoryear{Hopkins et al.}{2006}]{Hopkins
2006}Hopkins, P. F., 2006, ApJS, 163, 1H
\bibitem[\protect\citeauthoryear{Jeong et al.}{2009}]{Jeong
2009}Jeong, H et al., 2009, MNRAS, 398, 2028
\bibitem[\protect\citeauthoryear{Kauffmann et al.}{2003}]{Kauffmann 2003}Kauffmann G., White S. D. M., Heckman T. M. et al., 2003, MNRAS, 341, 54
\bibitem[\protect\citeauthoryear{Kennicutt.}{1998}]{Kennicutt 1998}Kennicutt, R. C. Jr, 1998, ARA\&A, 36, 189
\bibitem[\protect\citeauthoryear{Kormendy \& Kennicutt}{2004}]{Kormendy
2004}Kormendy, J. \& Kennicutt, R. C. Jr, 2004, ARA\&A, 42, 603
\bibitem[\protect\citeauthoryear{Lintott}{2008}]{Lintott
2008}Lintott, C. J., Schawinski, K., Slosar, A., et al., 2008, MNRAS, 389, 1179.
\bibitem[\protect\citeauthoryear{Lintott}{2011}]{Lintott
2011}Lintott, C. J., Schawinski, K., Bamford, S, et al., 2011, MNRAS, 410, 166.

\bibitem[\protect\citeauthoryear{Maraston et al.}{1998}]{Maraston 1998}Maraston,C, 1998, MNRAS, 300, 872.
\bibitem[\protect\citeauthoryear{Maraston et al.}{2005}]{Maraston 2005}Maraston,C, 2005, MNRAS, 362, 799.
\bibitem[\protect\citeauthoryear{Martin et al.}{2005}]{Martin 2005}Martin, D. C., 2005, ApJ, 619, L1
\bibitem[\protect\citeauthoryear{Masters et al.}{2011}]{Masters 2011}Masters, K. L., 2011,MNRAS, 411, 2026M
\bibitem[\protect\citeauthoryear{Mendez et al.}{2011}]{Mendez 2011}Mendez, A.
J., et al. 2011, ApJ, 736, 110
\bibitem[\protect\citeauthoryear{Nair \& Abraham}{2010}]{Nair 2010}Nair, P. B, \& Abraham, P. G. ,2010, ApJ, 714, 260
\bibitem[\protect\citeauthoryear{Oh et al.}{2011}]{Oh
2011}Oh,K., Sarzi, M., Schawinski, K., Yi, S. K., 2011, ApJS, 195, 13

\bibitem[\protect\citeauthoryear{Pan et al.}{2013}]{Pan 2013}Pan, Z. Z.,Kong, X \& Fan, L , 2013, ApJ, 776, 14

\bibitem[\protect\citeauthoryear{Peng et al.}{2010}]{Peng 2010}Peng Y. J., et al., 2010, ApJ, 721, 193

\bibitem[\protect\citeauthoryear{Salim et al.}{2005}]{Salim
2005}Salim, S. et al., 2005, ApJ, 619, L39

\bibitem[\protect\citeauthoryear{Suh et al.}{2010}]{Suh 2010}Suh, H. et al. 2010, ApJS, 187, 374
\bibitem[\protect\citeauthoryear{Schawinski et al.}{2010}]{Schawinski 2010}Schawinski. et al. 2010, ApJ, 714L, 108S
\bibitem[\protect\citeauthoryear{Schawinski et al.}{2014}]{Schawinski 2014}Schawinski. et al. 2014, MNRAS, 440, 889S
\bibitem[\protect\citeauthoryear{Scoville et al.}{2007}]{Scoville
2007}Scoville, N., Aussel, H., Brusa, M., et al., ApJS, 172, 1
\bibitem[\protect\citeauthoryear{Schlegel, Finkbeiner \& Davis}{1998}]{Schlegel
1998}Schlegel, D. J., Finkbeiner, D. P., \& Davis, M., 1998, ApJ, 500, 525S

\bibitem[\protect\citeauthoryear{Springel et al.}{2005}]{Springel
2005}Springel, V., Di, Matteo T., \& Hernquist, L., 2005, ApJ, 620, L79
\bibitem[\protect\citeauthoryear{Stark et al.}{2013}]{Stark 2013}Stark, D., et al., 2013, ApJ, 769, 82

\bibitem[\protect\citeauthoryear{Strateva et al.}{2001}]{Strateva
2001}Strateva, I. V et al., 2001, AJ, 122, 1861
\bibitem[\protect\citeauthoryear{Nair
& Abraham.}{2010}]{Nari 2010}Nair, P. B \& Abraham, R. G. ApJ, 714L, 260
\bibitem[\protect\citeauthoryear{Wang et al.}{2010}]{Wang 2010}Wang
J., et al., 2010, MNRAS, 401, 433

\bibitem[\protect\citeauthoryear{Wang et al.}{2011}]{Wang 2011}Wang
J., et al., 2011, MNRAS, 412, 1081W
\bibitem[\protect\citeauthoryear{Wang et al.}{2012}]{Wang 2012}Wang
J., et al., 2012, MNRAS, 423, 3486W
\bibitem[\protect\citeauthoryear{Wyder et al.}{2007}]{Wyder 2007}Wyder
T, K., et al., 2007, ApJS, 173, 293
\bibitem[\protect\citeauthoryear{Yang et al.}{2008}]{Yang 2008}Yang
Y. J, et al., 2008, ApJ, 688, 945
\bibitem[\protect\citeauthoryear{York et al.}{2000}]{York 2000}York
D. G, et al., 2000, AJ, 120, 1579

\end{thebibliography}
\end{document}